\documentclass[preprint,12pt]{elsarticle}

\usepackage{amssymb}
\usepackage{amsmath,amsfonts}
\usepackage{url}
\usepackage[dvipsnames]{xcolor}
\usepackage{textcomp}
\usepackage{manyfoot}
\usepackage{listings}
\usepackage{booktabs}
\usepackage{multirow}
\usepackage{float}

\journal{Information and Software Technology}

\graphicspath{{./images/}}

\begin{document}

\begin{frontmatter}

\title{Towards Understanding the Impact of Code Modifications on Software Quality Metrics}

\author{Thomas Karanikiotis\corref{cor1}}
\ead{karanikio@ece.auth.gr}
\author{Andreas L. Symeonidis}
\ead{symeonid@ece.auth.gr}
\cortext[cor1]{Corresponding author}

\affiliation{organization={Electrical and Computer Engineering Department},
            addressline={Aristotle University of Thessaloniki}, 
            city={Thessaloniki},
            country={Greece}}

\begin{abstract}

\textbf{Context:} In the realm of software development, maintaining high software quality is a persistent challenge. However, this challenge is often impeded by the lack of comprehensive understanding of how specific code modifications influence quality metrics.

\textbf{Objective:} This study ventures to bridge this gap through an approach that aspires to assess and interpret the impact of code modifications. The underlying hypothesis posits that code modifications inducing similar changes in software quality metrics can be grouped into distinct clusters, which can be effectively described using an AI language model, thus providing a simple understanding of code changes and their quality implications.

\textbf{Method:} To validate this hypothesis, we built and analyzed a dataset from popular GitHub repositories, segmented into individual code modifications. Each project was evaluated against software quality metrics pre and post-application. Machine learning techniques were utilized to cluster these modifications based on the induced changes in the metrics. Simultaneously, an AI language model was employed to generate descriptions of each modification's function.

\textbf{Results:} The results reveal distinct clusters of code modifications, each accompanied by a concise description, revealing their collective impact on software quality metrics.

\textbf{Conclusions:} The findings suggest that this research is a significant step towards a comprehensive understanding of the complex relationship between code changes and software quality, which has the potential to transform software maintenance strategies and enable the development of more accurate quality prediction models.
\end{abstract}



\end{frontmatter}

\section{Introduction}\label{sec:introduction}

\subsection{Background and Significance}

In an increasingly digitized world, software has evolved from being a mere utility to an integral part of our daily lives. It has infiltrated major business sectors, including finance, healthcare, education, and transportation, dictating the importance of maintaining high software quality \citep{Boehm2001}. It is no longer just about producing code, but ensuring that the code meets business requirements and standards of quality \citep{2013ISOI}. As a result, software quality research has experienced significant growth in recent years.

This growth is evident in the increasing focus on software product quality metrics \citep{9336003}, the application of machine learning techniques to predict software maintainability \citep{Gopal_Ms2019, su132212848, ISTmaintainabilityPaper} and improve software quality \citep{10.1145/1370175.1370234, 10.1145/1062455.1062614, 10.1145/1137702.1137703, icsoft2022karanikiotisbook}. These efforts reflect the recognition of the importance of software quality and the need for continuous improvement in this area.

Open-source software, a significant discipline/movement of this burgeoning field, allows anyone to access, modify, and share code \citep{10.1145/2145204.2145396}. GitHub\footnote{\url{https://github.com}}, as one of the largest code hosting platforms globally, with millions of developers contributing to diverse open-source projects \citep{7194588}, provides a wealth of data for researchers \citep{10.1145/2597073.2597126}. Commit history on GitHub serves as detailed records of a project's evolution over time, containing code changes and valuable metadata \citep{7886911}. These code modifications within commits can range from minor typo corrections to substantial patches affecting entire software components. We argue that studying these modifications unveils patterns and insights on how code evolves and how different changes can impact software quality \citep{karanikiotisICSOFT2020}. This exploration of modifications and their effect on software quality metrics represents an exciting, yet relatively unexplored area in software engineering research.

Maintainability, a critical quality characteristic, is defined as the effectiveness and efficiency with which a product or system can be modified by its intended maintainers \citep{seaman2001software}. It holds immense importance, as maintenance-related tasks, such as code refactoring, often consume a significant portion of software project efforts. The need for maintainable software is underscored by standards like ISO/IEC 25010:2011  \citep{2013ISOI}.

This research aims to explore code modifications' impact on software quality metrics, contributing to a thorough understanding of software quality in the context of digitalization and open-source collaboration. By querying GitHub repositories, analyzing code modifications, calculating static analysis metrics, and applying clustering techniques, we seek to uncover the complex relationship between code changes and software quality, shedding light on the multifaceted nature of maintainability in the ever-evolving landscape of software development. Furthermore, we harness the power of advanced AI models, such as GPT-4, to assist in summarizing code modifications, providing additional context for our findings.

In the following sections, we will delve into our methodology, detailing how we collect and analyze code modifications and their impact on software quality metrics. Through this research, we aim to contribute valuable insights that can inform best practices in software development, ultimately fostering the creation of more maintainable software components.

\subsection{Purpose of the Study}

Considering this context, the primary purpose of this study is to delve deeper into the relationship between different code modifications and their subsequent impact on software quality metrics. We intend to explore this relationship in a comprehensive manner by employing a multi-step approach.

Firstly, we plan to collect a substantial dataset from a selection of popular GitHub repositories, with a focus on the commits and the code modifications contained within them. The diversity and richness of these repositories should provide a wide range of code changes for analysis.

The second step involves calculating static software quality metrics both before and after each modification. Static code analysis metrics offer quantifiable measures of various aspects of code quality, such as complexity, maintainability, or potential vulnerabilities, thus making them a suitable tool to gauge the impact of code modifications.

The third step introduces a novel element to the analysis by using an AI model to generate understandable summaries of these code modifications. The aim of this step is to bridge the gap between complex code changes and human interpretation, providing a tool that simplifies the task of comprehending the nature of these modifications.

Finally, we aim to perform a clustering analysis of these modifications based on their impact on software quality metrics. The goal of this clustering is to group together modifications that have similar effects, thereby identifying patterns and commonalities among code changes. Through this process, we hope to contribute to a better understanding of how code evolves and how its quality can be maintained or improved over time.

By exploring these relationships within repositories, this study seeks to shed new light on the dynamics of software quality evolution in the context of code modifications, offering valuable insights for software developers, project managers, researchers, and all those involved in the realm of software development and maintenance.

\subsection{Structure of the Paper}

The rest of this paper is organized as follows:

\begin{itemize}

\item In section \ref{sec:literature-review}, we provide a comprehensive overview of existing studies in the field, with a focus on software quality metrics, methods for evaluating code quality, and the analysis of code modifications and evolution. The section also identifies gaps in the current literature that this study aims to address.

\item In section \ref{sec:methodology}, we explain the detailed steps taken in our research, namely the collection and pre-processing of data, the calculation of software quality metrics, the summarization of code modifications using an AI model, the application of a clustering algorithm, and the subsequent evaluation and interpretation of the clusters.

\item In section \ref{sec:results}, we provide a comprehensive overview of our research findings. This includes a detailed examination of the computed quality metrics, insightful AI-generated summaries highlighting the essence of each modification, and a thorough presentation of the outcomes derived from the clustering process. Our analysis includes a range of examples, showcasing modifications with varying impacts on software quality metrics, from those with undesirable effects to those that contribute positively to software quality.

\item In section \ref{sec:discussion}, we present insight to our results, while stressing implications and potential applications.

\item Finally, in section \ref{sec:conclusions-and-future-work}, we offer a summary of the research and provide insights for further research.

\end{itemize}

\section{Literature Review}\label{sec:literature-review}

The literature on software quality metrics and their relationship to code modifications is both extensive and diverse. This body of research has sought to understand, quantify, and improve software quality by examining various factors, including the characteristics of the codebase, the nature of code modifications, and the use of artificial intelligence techniques in software engineering. This section provides a review of the most relevant studies in this area, establishing the context for the present research and highlighting the gaps that our study aims to address.

In the realm of software quality metrics, significant research has been conducted on complexity, maintainability, reusability, and readability, amongst other aspects. Various tools and techniques have been developed to quantify these metrics, providing valuable insights into the quality of software code. These metrics serve as the foundation for assessing the impact of code modifications on software quality, a relationship that has been explored in several studies.

In their research, Stroggylos and Spinellis \citep{4273477} studied the logs of version control systems in prevalent open-source software systems to understand the impact of refactoring, a standard type of code modification, on software metrics. They discovered that the expected quality improvement through refactoring was not uniformly reflected in the metrics. This indicates that the correlation between code modifications and software quality metrics is more intricate than previously understood.

Considering the importance of test code quality, Athanasiou et al. \citep{6862882} introduced a model for evaluating it by combining source code metrics that denote completeness, effectiveness, and maintainability. This underlines the necessity of ensuring high-quality test code as it directly impacts issue resolution performance.

Meirelles et al. \citep{5631691} explored how source code metrics relate to the appeal of free software projects. They concluded that certain quantifiable attributes of the project's source code influence the decision to contribute to and use the software, emphasizing the significance of monitoring source code quality.

In the context of code review, Kononenko et al. \citep{7332457} investigated the factors affecting the quality of the code review process. They found a connection between the quality of code reviews and both personal metrics and participation metrics, thus signifying the role of thorough code reviews in maintaining or improving software quality.

Trautsch et al. \citep{Trautsch2023} undertook a commit-level study to understand the effect of changes in static metric value and static analysis warning on source code quality enhancements. They established that the size and nature of commits can significantly impact static metrics and consequently, software quality.

Investigating ownership metrics, Foucault et al. \citep{FOUCAULT2015102} verified a connection between code ownership and software quality. However, the importance of ownership metrics in relation to other metrics remains an open question.

Co-change metrics have been used to predict software bugs \citep{6606741}, and understanding the dispersion characteristics of co-changes can assist developers in maintaining software quality.

Additionally, the role of static code analysis tools in shaping developers' understanding and decision-making has been emphasized in \citep{10.1109/ICSE43902.2021.00055}. These tools aid developers in making well-informed decisions while modifying code to uphold and enhance software quality.

In our prior study \citep{su132212848}, we emphasized the importance of proactive measures for maintaining software quality amidst agile development's rapid changes and releases. Our approach, based on code hosting platform data, identified non-maintainable software classes and highlighted the value of tracking evolving static analysis metrics. This aligns with our current research goal of analyzing how code modifications impact software quality metrics.

Collectively, prior studies have delved into specific facets of software quality metrics and the impact of individual code modification types. While these investigations are invaluable, they tend to view these elements in isolation, lacking a holistic perspective that considers the complex interplay between different types of code modifications and their collective effect on software quality metrics. Our research takes a step beyond the fragmented approach of previous studies by acknowledging that software quality metrics often have competing interests. For instance, while one metric may improve due to a certain code modification, another may decline.

In this work, we set out to fill this gap. We undertake a comprehensive approach that considers multiple types of code modifications, examining their impacts on a diverse array of software quality metrics. Our methodology allows us to explore the intricate relationship between various categories of code modifications and their impact on software quality. This exploration potentially reveals new insights into the effects of code modifications on software quality, providing a richer and more complex understanding of these dynamics.

\section{Materials and Methods}\label{sec:methodology}

This section outlines the comprehensive methodology and architecture employed in our research, which encompasses the analysis of data from GitHub repositories. Our primary objective is to investigate the impact of diverse code modifications on software quality metrics. To achieve this, our study integrates static code analysis, language processing using AI language models, and clustering analysis.

Our methodology comprises four primary phases: data collection, preprocessing, pre-/post-commit quality metrics calculation, and interpretation of results through clustering of code modifications. We initiated the process by systematically gathering commit data from GitHub, encompassing a wide spectrum of repositories. Each commit was than meticulously split into individual modifications for a fine-grained examination.

For every modification, we conducted static code analysis both before and after the change, generating metrics that shed light on the alteration's influence on code quality. To enhance the comprehensibility of this analysis, we harnessed the capabilities of an AI language model to generate human-readable descriptions of these modifications.

Lastly, a clustering algorithm was deployed to categorize the modifications based on their impact on software quality metrics. These clusters were subsequently assessed, and the generated summaries were utilized to comprehend which modifications exhibited similar effects on quality metrics.

\begin{figure}[ht]
\centering
\includegraphics[width=\textwidth]{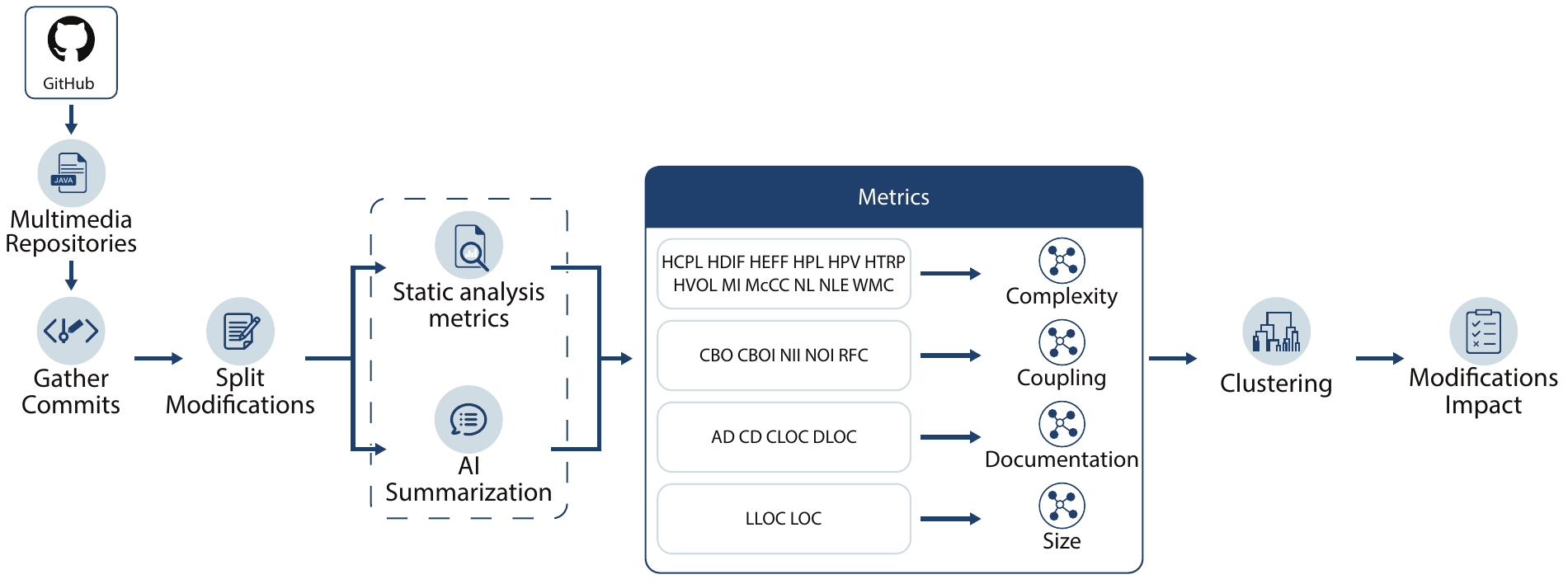}
\caption{The architecture of our approach.}\label{fig:architecture}
\end{figure}

The accompanying architectural diagram in Figure \ref{fig:architecture} provides a visual representation of our methodology, illustrating the sequential data flow from collecting commits to deriving insights from the clustered modifications. Each facet of this architecture contributes to our primary goal: comprehending how diverse code modifications influence software quality metrics. Further elaboration on each phase is presented in the subsequent sections of the methodology.

\subsection{Gathering Commits from GitHub Repositories}

The initial step in our methodology involves the systematic collection of commits from GitHub repositories. Given the immense number of GitHub repositories, we adopted a pragmatic approach by including 700 repositories that included almost 80,000 commits. This comprehensive dataset ensures a thorough exploration of code modifications and their impact on software quality metrics.

To determine their popularity and engagement within the developer community, we considered both star and fork counts. As it is well-known, stars reflect the degree of interest in a project, while forks indicate how many developers have utilized the repository as a foundation for their own projects. This combination of metrics provides valuable insights into community-driven engagement. In selecting repositories for our study, we specifically targeted those with significant traction, focusing on the top starred and forked repositories. We set a threshold of more than 1000 stars and 1000 forks to ensure that the repositories chosen had substantial community interest and activity. This rigorous selection criteria aimed to provide us with repositories that are not only widely recognized but also actively used and contributed to by developers. By focusing on repositories with high star and fork counts, we aimed to capture a diverse range of projects with varying levels of community engagement, thereby enriching the depth and breadth of our dataset for analysis.

In our increasingly digital world, we delve into the multimedia domain and its connection to cultural heritage. UNESCO classifies cultural heritage into tangible and intangible categories, ranging from paintings to oral traditions \citep{unesco2003convention}. Preserving this heritage requires modern multimedia tools. Software development trends also hold cultural value, mirroring artifacts like computer science fiction movies, games, and music. Much like preserving tangible heritage, software development must ensure long-term viability and maintainability \citep{su13031079}. Our choice of the multimedia domain serves as a comprehensive case study, demonstrating that our research's insights are not domain-specific and can be applied more broadly.

Some of the most popular multimedia-related repositories from our dataset are depicted in Table \ref{tab:repositories}.

\begin{table}[ht]
\caption{The 10 most popular repositories from our collected dataset.}\label{tab:repositories}
\centering
\begin{tabular}{lcc}
\toprule
\textbf{Repository} & \textbf{Stars} & \textbf{Forks}\\
\midrule
CarGuo/GSYVideoPlayer & 18,991 & 4,086\\
LuckSiege/PictureSelector & 12,697 & 2,935\\
GoogleCloudPlatform/java-docs-samples & 1,608 & 2,827\\
TeamNewPipe/NewPipe & 25,292 & 2,764\\
react-native-video/react-native-video & 6,686 & 2,758\\
lipangit/JiaoZiVideoPlayer & 10,458 & 2,437\\
zhihu/Matisse & 12,444 & 2,080\\
RameshMF/spring-boot-tutorial & 1,452 & 1,677\\
bytedeco/javacv & 6,896 & 1,559\\
648540858/wvp-GB28181-pro & 3,650 & 1,146\\
\bottomrule
\end{tabular}
\end{table}

For each repository selected, we cloned it to a local directory and employed PyDriller\footnote{\url{https://pypi.org/project/PyDriller/}} to traverse the commit history. For each commit, we compiled a list of modified files. We focused exclusively on files with a \textit{.java} extension, signifying Java files. This selection has been made without loss of generality, as it serves only as a case study. Our approach is completely language-agnostic and allows the incorporation of files from other programming languages. For each qualifying modified file, we extracted and stored critical information, including:

\begin{itemize}

\item{\textbf{Code diff}: Representing the changes made to the file in that specific commit.}

\item{\textbf{SHA}: A unique identifier for the commit.}

\item{\textbf{Filename}: The name of the modified file.}

\item{\textbf{Commit message}: The developer's message accompanying the commit.}

\item{\textbf{Code before}: The state of the file's code prior to the commit.}

\item{\textbf{Code after}: The state of the file's code following the commit.}

\end{itemize}

This data was meticulously saved for each file, forming a comprehensive dataset that encapsulates file changes along with associated metadata. Table \ref{tab:commits} provides an overview of the collected commit data.

\begin{table}[ht]
\caption{The collected commits.}\label{tab:commits}
\centering
\begin{tabular}{lcc}
\toprule
\textbf{Metric} & \textbf{Total Value} & \textbf{Average Value per Repository}\\
\midrule
Commits & 79,178 & 118\\
Modified Java Files & 368,631 & 550\\
\bottomrule
\end{tabular}
\end{table}

\subsection{Splitting Commits into Modifications}

Having collected the commits from our selected repositories, we proceeded to the next phase of our methodology: splitting these commits into individual modifications. This stage is crucial because it enables us to gain an in-depth understanding of the specific changes that each commit introduces to the codebase. In the context of this study, a modification is defined as any change made to a code segment that results in a transformation of its structure or functionality. This can range from a single-line change, such as the modification of a variable's value, to the addition or removal of whole methods or classes.

In order to break down the commits into modifications, we analyzed the code diffs provided by GitHub for each commit. A diff in this context is a formatted display of the changes made in a commit. It shows what was added or removed from a file's content, line by line. To split the modifications within each file into meaningful snippets, we used a syntactically-aware approach. This means that we took into account the syntactical structure of the Java code to ensure that the modifications we extract are self-contained and syntactically correct.

For instance, if a change was made in the import declarations or variable declarations, we made sure to include the relevant portions of the code to maintain the integrity of the modifications. This approach allowed us to minimize the possibility of extracting code snippets that couldn't be compiled due to missing context or information.

Additionally, we ensured that the commits were split based on the part of the code altered, as indicated in the GitHub diff. This approach allows for a granular analysis of the code changes, with different areas in the diff representing different modifications. Each modification was then saved separately, along with its associated metadata (i.e., the commit SHA, filename, and commit message). The result was a comprehensive dataset of modifications, each representing a single change in the codebase.

Table \ref{tab:modifications} provides information about the modifications extracted by the above mentioned approach.

\begin{table}[ht]
\caption{The collected modifications.}\label{tab:modifications}
\centering
\begin{tabular}{lc}
\toprule
\textbf{Metric} & \textbf{Value}\\
\midrule
Total Number of Modifications & 1,758,931\\
Average Number of Modifications per Repository & 2,625\\
\bottomrule
\end{tabular}
\end{table}

\subsection{Static Quality Metrics}

Once we had separated the commits into individual modifications, we needed to evaluate the impact of each change on the codebase's quality. To achieve this, we calculated the static quality metrics for the code before and after each modification.

Following the ISO/IEC 25010:2011 standard \citep{2013ISOI}, the quality degree of a software component is composed by various sub-characteristics, which are described at length in \citep{bass2003software}, and comprises eight quality characteristics: \textit{Functional Suitability}, \textit{Usability}, \textit{Maintainability}, \textit{Portability}, \textit{Reliability}, \textit{Performance} and \textit{Efficiency}, \textit{Security} and \textit{Compatibility}, out of which the first four are often evaluated by developers in an intuitive manner. 

To quantify these external product quality categories, a set of metrics derived from static analysis and computation levels can be employed. For instance, in terms of Complexity, metrics such as Halstead Calculated Program Length (HCPL), McCabe's Cyclomatic Complexity (McCC), and Weighted Methods per Class (WMC) are utilized to gauge the intricacy of the software structure and its potential impact on maintainability and reliability. Coupling metrics like Coupling Between Object classes (CBO), Number of Incoming Invocations (NII), and Number of Outgoing Invocations (NOI) help assess the level of interdependency among software components, which influences both performance efficiency and maintainability. Additionally, documentation-related metrics such as Comment Density (CD) and Documentation Lines of Code (DLOC) provide insights into the clarity and comprehensiveness of the codebase, contributing to usability and maintainability aspects. Finally, size-related metrics like Logical Lines of Code (LLOC) and Lines of Code (LOC) offer quantitative measures of the software's scale, which is crucial for evaluating its portability and performance efficiency across different environments. Dimaridou et al. \citep{Dimaridou2017} have depicted the correlation between the quality characteristics and the quality metrics. By employing these metrics across various properties, software quality can be more comprehensively evaluated, encompassing both internal code characteristics and their external impact on product quality \citep{su132212848}.

The metrics we selected are categorized as follows:

\begin{itemize}

\item{Complexity Metrics: Complexity metrics in software engineering, such as cyclomatic complexity and depth of inheritance, provide quantitative measures to assess the intricacy of code, serving as indicators of the potential challenges in understanding, maintaining, or modifying the software.}

\item{Coupling Metrics: Coupling metrics quantify the degree of interdependence between software components or modules. High coupling can indicate increased complexity and potential maintenance challenges.}

\item{Documentation Metrics: Documentation in the context of software engineering refers to the written text and diagrams that explain the design, functionality, and use of the software. Well-documented code is easier to understand, maintain, and modify.}

\item{Size Metrics: Size metrics gauge the extent of codebase growth or reduction resulting from modifications. Metrics such as lines of code added or deleted provide insights into the scale of code changes.}

\end{itemize}

Table \ref{tab:quality_metrics} below provides a full reference of the computed metrics along with their computation level (method or class). An upward-pointing arrow signifies that a positive alteration in the metric value contributes positively to the overall quality. Conversely, a downward-pointing arrow indicates that a negative adjustment in the metric value is necessary to enhance the quality score of the code. Lastly, a dash indicates that the metric is not applicable at the current computational level.

\begin{table}[ht]
\footnotesize
\caption{The calculated metrics per source code property. \textit{Doc} stands for documentation}\label{tab:quality_metrics}
\centering
\begin{tabular}{cclcc}
\toprule
 & \multicolumn{2}{c}{\textbf{Static Analysis Metrics}} & \multicolumn{2}{c}{\textbf{Computation Levels}}\\
\midrule
\textbf{Property} & \textbf{Name} & \textbf{Description} & \textbf{Method} &\textbf{Class}\\
\midrule
\multirow{12}{*}{\textbf{\rotatebox[origin=c]{90}{Complexity}}} & HCPL & Halstead Calculated Program Length & $\searrow$ & -\\
 & HDIF & Halstead Difficulty & $\searrow$ & -\\
 & HEFF & Halstead Effort & $\searrow$ & -\\
 & HPL & Halstead Program Length & $\searrow$ & -\\
 & HPV & Halstead Program Vocabulary & $\searrow$ & -\\
 & HTRP & Halstead Time Required to Program & $\searrow$ & -\\
 & HVOL & Halstead Volume & $\searrow$ & -\\
 & MI & Maintainability Index & $\nearrow$ & -\\
 & McCC & McCabe's Cyclomatic Complexity & $\searrow$ & -\\
 & NL & Nesting Level & $\searrow$ & $\searrow$\\
 & NLE & Nesting Level Else-If & $\searrow$ & $\searrow$\\
 & WMC & Weighted Methods per Class & - & $\searrow$\\
\midrule
\multirow{5}{*}{\textbf{\rotatebox[origin=c]{90}{Coupling}}} & CBO & Coupling Between Object classes & - & $\searrow$\\
 & CBOI & CBO Inverse & - & $\searrow$\\
 & NII & Number of Incoming Invocations & $\searrow$ & $\searrow$\\
 & NOI & Number of Outgoing Invocations & $\searrow$ & $\searrow$\\
 & RFC & Response set For Class & - & $\searrow$\\
\midrule
\multirow{4}{*}{\textbf{\rotatebox[origin=c]{90}{Doc}}} & AD & API Documentation & - & $\nearrow$\\
 & CD & Comment Density & $\nearrow$ & $\nearrow$\\
 & CLOC & Comment Lines of Code & $\nearrow$ & $\nearrow$\\
 & DLOC & Documentation Lines of Code & $\nearrow$ & $\nearrow$\\
\midrule
\multirow{2}{*}{\textbf{\rotatebox[origin=c]{90}{Size}}} & LLOC & Logical Lines of Code & $\searrow$ & $\searrow$\\
 & LOC & Lines of Code & $\searrow$ & $\searrow$\\
\bottomrule
\end{tabular}
\end{table}

To calculate these metrics, we utilized SourceMeter\footnote{\url{https://sourcemeter.com}}, an innovative tool designed for in-depth source code analysis. SourceMeter enabled us to automatically measure a variety of software quality metrics directly from our collected Java files. For each modification, we ran SourceMeter on the code both before and after the modification was applied. 

To assess the impact of code modifications on software quality metrics, we implemented a meticulous methodology. After isolating each commit into individual modifications, we embarked on a detailed evaluation of the metrics associated with the code's methods and classes both before and after each modification.

In this process, we first computed the static quality metrics for all methods and classes within the file undergoing modification. This comprehensive analysis provided us with baseline measurements of the code's quality characteristics before any changes were made. Subsequently, we identified the specific methods and classes that were directly affected by the modification, focusing on their metrics before and after the alteration.

To gauge the modification's impact accurately, we calculated the percentage difference in metrics for each affected method and class. This involved comparing the metric values before and after the modification, expressing the change as a percentage relative to the initial metric value. By employing this granular approach, we obtained precise insights into how individual methods and classes were influenced by the code modification.

Finally, to calculate the modification's impact on software quality metrics, we computed the average percentage difference for both methods and classes. This cumulative percentage difference served as a single, comprehensive indicator of the modification's influence on the codebase's quality. This methodology allowed us to offer an exact understanding of how code modifications ripple through software components, affecting their quality metrics in a cohesive manner.

An important part of our methodology involves filtering out data that exhibited zero differences in all metrics used within each cluster. Notably, a significant proportion of code modifications fell into this category, reflecting their relatively small and straightforward nature. These modifications introduced minimal changes to the codebase, resulting in negligible variations in software quality metrics. While these data points were part of our collected set, they did not contribute meaningful information to our analysis, as they represented instances where no substantial change occurred from a software quality metrics' perspective.

Including such data in our clusters would have potentially skewed our results by diluting the impact of modifications that genuinely influenced software quality metrics. Thus, to ensure a focused and precise analysis of the relationship between code modifications and quality metrics, we excluded these instances from our clustering process. This careful filtering allowed us to concentrate on the subset of code modifications that had a discernible effect on software quality metrics, leaving us with \textbf{532} meaningful modifications for our analysis. This approach enhanced the reliability and accuracy of our findings.

\subsection{Using AI for Summarizing Modifications}

To enhance our understanding of the results and facilitate data interpretation, we employed an artificial intelligence model for generating summaries of code modifications. It is essential to note that this step serves as a supplementary aid to help digest the research findings, and it is not a primary focus of our study.

Given the substantial volume and complexity of the code modifications, we leveraged OpenAI's state-of-the-art language model, GPT-4\footnote{\url{https://openai.com/research/gpt-4}}. This advanced AI model, equipped with natural language processing capabilities, excels at producing concise and coherent text summaries based on input data.

Our summarization process follows two distinct steps. In the first step, we query the AI model for a summary  for each code modification. Next  we distill the summary into a simpler, more generalized description. The AI-generated summaries play a pivotal role in our dataset, bridging the gap between raw data and human understanding. They provide essential support in subsequent stages, including clustering and interpretation.

To illustrate our approach, consider the following code modification:

\begin{figure}[ht]
\par\noindent\rule{\linewidth}{0.4pt}
\begin{lstlisting}[
           language=Java,
           showspaces=false,
           basicstyle=\sffamily\footnotesize,
        ]
- private void fetchData(){
+ public void fetchData(){
\end{lstlisting}
\par\noindent\rule{\linewidth}{0.4pt}
\caption{Example code modification from our dataset.}
\label{fig:example-summary}
\end{figure}

In this instance, a private method "\textit{fetchData}" has been modified to become public. When this code modification is input into the GPT-4 model, we obtain the following summarization:

\begin{center}
\emph{"Access modifier changed from private to public."}
\end{center}

In the second step of the summary generation, the simpler summary is the following:

\begin{center}
\emph{"Changed access modifier."}
\end{center}

This concise description aptly captures the essence of the modification. Throughout our analysis, it is these AI-generated summaries that we leveraged to provide an interpretable representation of each modification. This process greatly aided our clustering and interpretation phases, enabling a deeper understanding of the relationship between code modifications and quality metrics.

\subsection{Clustering Based on the Impact on Quality Metrics}

In order to group the code modifications based on their impact on software quality, we employed a clustering technique. Specifically, we employed the K-means algorithm, a typical, yet well-established clustering method known for its ease of application and efficiency.

The K-means algorithm partitions data into k distinct clusters based on distance to the centroid of the clusters. For our case, we fed the algorithm the percentage differences of the metrics corresponding to each code modification. This enabled the algorithm to group the modifications according to the similarity of their impacts on the quality metrics.

In order to perform the clustering analysis, we sought to determine the optimal number of clusters. To do this, we tested a range of clusters, depending on the available data, and evaluated each configuration based on the silhouette score. The calculation of the silhouette coefficient in \ref{eq:sil} uses the mean intra-cluster distance and the minimum nearest cluster distance, as shown in equations \eqref{eq:intracluster} and \eqref{eq:nearest_cluster}.

\begin{equation}\label{eq:intracluster}
a(i) = \frac{1}{|c_i| - 1} \sum_{\substack{j \in C_i\\
                  j \neq i}} 
                  {d(i, j)}
\end{equation}

\begin{equation}\label{eq:nearest_cluster}
b(i) = \min_{k \neq i} \frac{1}{|c_k|} \sum_{j \in C_k} {d(i, j)}
\end{equation}

\begin{equation}\label{eq:sil}
s(i) = \frac{b(i) - a(i)}{max(a(i), b(i))}
\end{equation}

\subsection{Evaluation of Clusters}

Following the generation of the clusters, our focus shifted to their interpretation, in order to understand their significance in terms of the impact of code modifications on different facets of software quality. Our evaluation process employed a selection criterion based on silhouette scores, the number of modifications per cluster, and the distribution of modifications across different repositories.

Firstly, we paid careful attention to the quality of the clusters, underpinned by the silhouette scores, which had been instrumental in determining the optimal number of clusters. Clusters were considered of high quality if their silhouette scores exceeded a predefined threshold, signifying well-separated and internally cohesive clusters. For this evaluation, we set the silhouette score threshold equal to the mean silhouette score of all the clusters, dropping the ones that do not appear to have strong internal cohesion.

The next criterion was the number of modifications within each cluster. To avoid working with clusters that had a very sparse representation of modifications, we established a minimum count of 5 modifications that a cluster must encapsulate. This decision was driven by our aim to derive meaningful and statistically significant insights from the clusters.

Lastly, to ensure the robustness and general applicability of our results, we introduced a condition related to the repository distribution of modifications. We only considered those clusters for further analysis where modifications were spread across more than one repository. This criterion was designed to avoid clusters representing local, file-specific changes, thereby ensuring that our analysis did not get skewed by the unique characteristics or contexts of individual repositories.

By establishing these strict selection criteria, we aimed to evaluate clusters that offered rich, generalized, and statistically robust insights into the types of code modifications that similarly impacted software quality metrics. This rigorous approach ensured the practical utility and reliability of our findings, making them invaluable for informed decision-making during code review and maintenance processes.

\section{Results}\label{sec:results}

Our core findings lie in the changes in quality metrics before and after the code modifications. This is followed by the outcomes of our clustering process using K-means and a detailed analysis of the resulting clusters. We present the evaluation results of these clusters and identify key insights, trends, and patterns observed.

\subsection{Code Modifications Summaries}

In this section, we will unveil the outcomes of the GPT-4 summarization process applied to code modifications. These summarizations offer a concise yet informative representation of the technical changes, enabling a more accessible assessment of their impact on software quality metrics.

The following figures present a random selection of code modifications, along with their AI-generated summaries given as captions. As depicted, the model was able to succinctly and accurately describe the essence of each modification. The code in Figure \ref{fig:summarizaton_example_2} corresponds to a modification that removes unnecessary and unused import statements, with the initial summary capturing the specific change with the summary \emph{"Remove unused imports of animation libraries"}. The second summary (\emph{"Remove unused imports"}) describes the same modification in a more simplified and generalized manner, capturing the fundamental essence of the change while omitting specific technical details. This simplification process helps us identify patterns and themes among code modifications, ultimately aiding in more meaningful clustering and interpretation of our research results.

\begin{figure}[ht]
\par\noindent\rule{\linewidth}{0.4pt}
\begin{lstlisting}[
           language=Java,
           showspaces=false,
           basicstyle=\sffamily\footnotesize,
        ]
-import android.animation.ObjectAnimator;
-import android.animation.ValueAnimator;
-import android.animation.ValueAnimator.AnimatorUpdateListener;
\end{lstlisting}
\par\noindent\rule{\linewidth}{0.4pt}
\caption{\textbf{Summary}: \textit{Remove unused imports of animation libraries.} \textbf{Simpler summary}: \textit{Remove unused imports.}}
\label{fig:summarizaton_example_2}
\end{figure}

\begin{figure}[ht]
\par\noindent\rule{\linewidth}{0.4pt}
\begin{lstlisting}[
           language=Java,
           showspaces=false,
           basicstyle=\sffamily\footnotesize,
        ]
-            mYValueSum += Math.abs(mYVals.get(i).getVal());
+            mYValueSum += Math.abs(mYVals.get(i).getSum());
\end{lstlisting}
\par\noindent\rule{\linewidth}{0.4pt}
\caption{\textbf{Summary}: \textit{Update method to get sum instead of single value.} \textbf{Simpler summary}: \textit{Changed method call.}}
\label{fig:summarizaton_example_3}
\end{figure}

\begin{figure}[ht]
\par\noindent\rule{\linewidth}{0.4pt}
\begin{lstlisting}[
           language=Java,
           showspaces=false,
           basicstyle=\sffamily\footnotesize,
        ]
+    /**
+     * Returns the data, additional information that
+     * this Entry represents, or
+     * null, if no data has been specified.
+     *
+     * @return
+     */
+    public Object getData() {
+        return mData;
+    }
+
+    /**
+     * Sets additional data this Entry should represents.
+     *
+     * @param data
+     */
+    public void setData(Object data) {
+        this.mData = data;
+    }
\end{lstlisting}
\par\noindent\rule{\linewidth}{0.4pt}
\caption{\textbf{Summary}: \textit{Add methods to get and set additional data for a given entry.} \textbf{Simpler summary}: \textit{Add new methods.}}
\label{fig:summarizaton_example_4}
\end{figure}

Furthermore, as their summaries indicate, the modification in Figure \ref{fig:summarizaton_example_3} updates the method used to retrieve a specific value, while the respective one in Figure \ref{fig:summarizaton_example_4} adds two more useful functions in the codebase. This multi-step summarization approach proved invaluable in providing concise yet informative representations of the code modifications, facilitating our subsequent analysis.

\begin{figure}[ht]
\centering
\includegraphics[width=\textwidth]{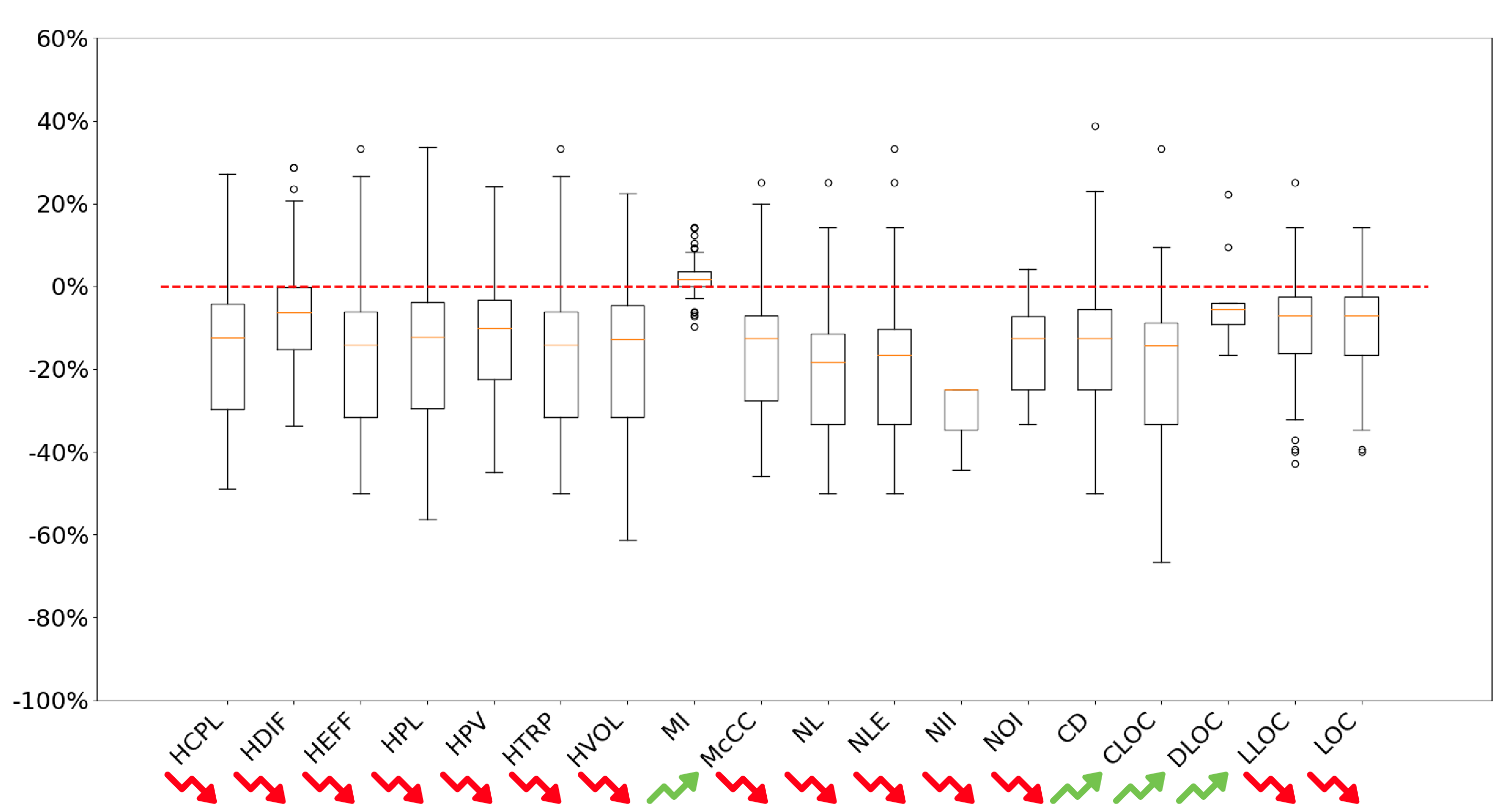}
\caption{The quality metrics on method level difference percentage distribution.}\label{fig:methods_metrics_boxplot}
\end{figure}

These summaries significantly eased the subsequent analysis stages, allowing a more efficient and comprehensive understanding of the modifications' impact on software quality metrics, without necessitating a deep dive into the specifics of the code changes.

\subsection{Quality Metrics}

One of the primary objectives of this research was to examine the impact of code modifications on various software quality metrics. These metrics, derived from the ISO/IEC 25010 model, span across different aspects of software quality, including complexity, coupling, documentation, and size.

\begin{figure}[ht]
\centering
\includegraphics[width=\textwidth]{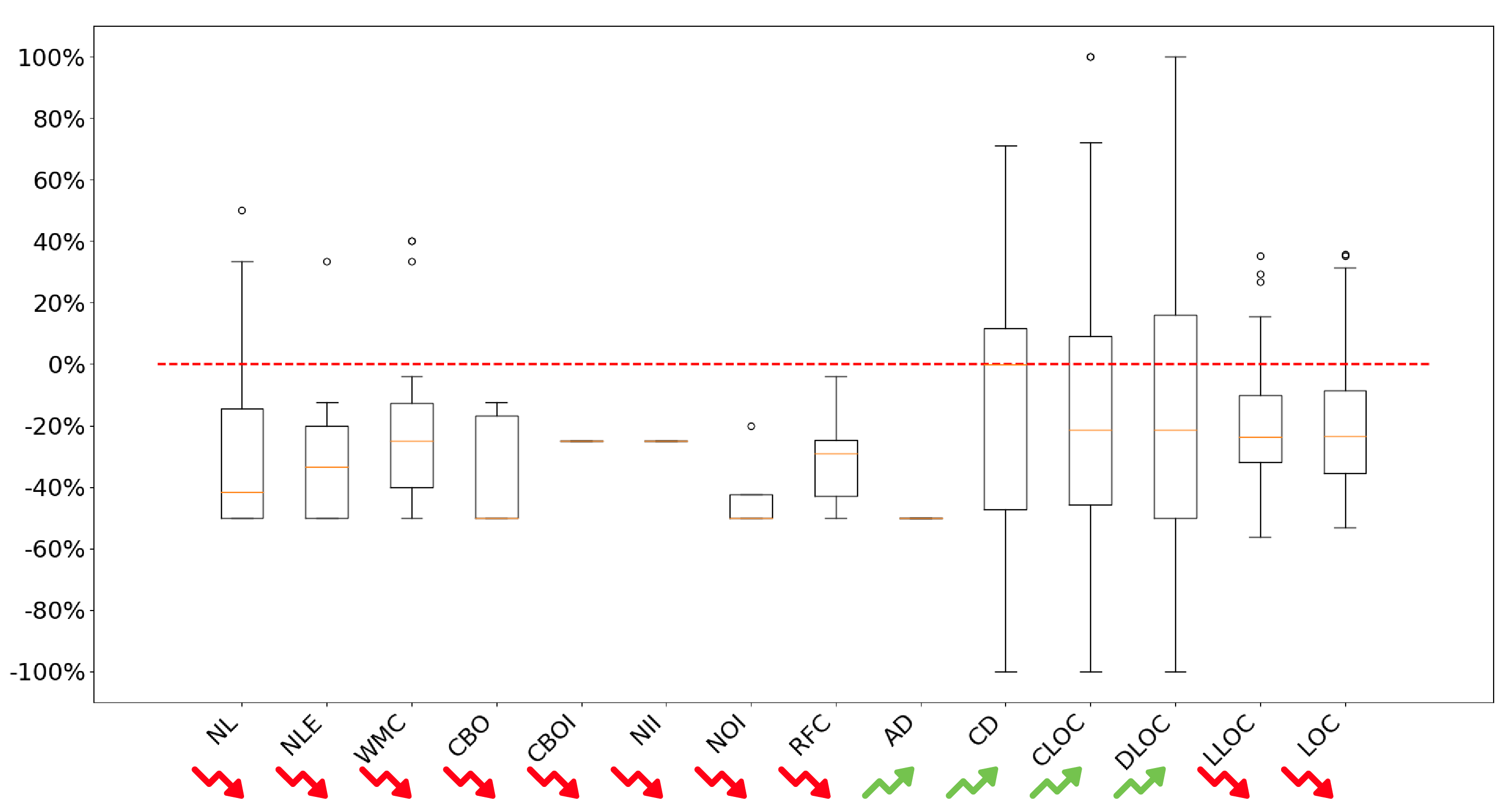}
\caption{The quality metrics on class level difference percentage distribution.}\label{fig:classes_metrics_boxplot}
\end{figure}

Figure \ref{fig:methods_metrics_boxplot} provides a high-level view of the changes in these metrics before and after code modifications were applied in a method level. The boxplots provide a total overview of the difference in the values of each metric as they were counted before and after the modification, on the methods that were altered by each modification. It should be noted again that we have excluded modifications with absolutely zero impact on each metric from our analysis as they provide no informative data; these modifications did not result in any measurable alterations in the respective quality metrics. The arrows below the name of each metric indicate the direction towards improving the overall quality score. For example, the overall quality score increases when the \textit{HCPL} metric decreases or the \textit{MI} metric increases. Figure \ref{fig:classes_metrics_boxplot} illustrates the respective boxplots for the metrics that are calculated on class level.

These boxplots provide a visual representation of the distribution and variations in software quality metrics both in method and class level, as defined by ISO/IEC 25010:2011. Analyzing these boxplots allows for a deeper understanding of how the modifications tend to impact the quality of the software. In the following bullet points, we delve into some key conclusions drawn from this analysis:

\begin{itemize}
    \item In the case of \emph{McCabe's Cyclomatic Complexity (McCC)}, the majority of the modifications appear a negative difference, which means that, usually, the modifications tend to decrease the cyclomatic complexity of the code.
    \item Intriguingly, the metrics \emph{Comment Density (CD)}, \emph{Comment Lines of Code (CLOC)} and \emph{Documentation Lines of Code (DLOC)} typically demonstrate a negative difference, indicating a decrease in these attributes post-modification. The decrease in these metrics could be due to a focus on optimizing the code structure and performance over documentation. By reducing the amount of comments and documentation, developers might inadvertently make the code less understandable, which can affect maintainability.
    \item Software metrics that affect the coupling of the codebase, like \emph{NII} and \emph{NOI} (both in method and in class level), as well as \emph{CBO}, \emph{CBOI} and \emph{RFC} in class level, show a negative difference only in few cases, while usually exhibit zero difference before and after the modification. This suggests that the coupling of the code alters only slightly during the modification process. This stability in coupling metrics indicates that the fundamental structural relationships among objects and classes remain relatively unchanged. This, in turn, suggests that the modification process tends to preserve the existing code coupling characteristics, which can have implications for maintainability and future development efforts.
    \item It is noteworthy that size metrics, specifically lines of code (\emph{LOC}) and logical lines of code (\emph{LLOC}), consistently exhibit negative differences, indicating a reduction after the modification process. This intriguing trend not only points to a decrease in code size but also suggests a reduction in overall complexity. The decrease in complexity metrics aligns with the observed reduction in size, indicating that software modifications often lead to a simplification of the codebase. This reduction in complexity can have several advantageous outcomes, such as enhanced maintainability, a lower potential for defects, and improved code comprehensibility. 
\end{itemize}

\subsection{Clustering Results}

Following the application of the K-means clustering algorithm to the percentage differences in quality metrics, we have obtained meaningful results that contribute significantly to our understanding of the impact of different code modifications on software quality. Figure \ref{fig:silhouette} below illustrates the silhouette scores obtained during the clustering analysis.

\begin{figure}[ht]
\centering
\includegraphics[width=\textwidth]{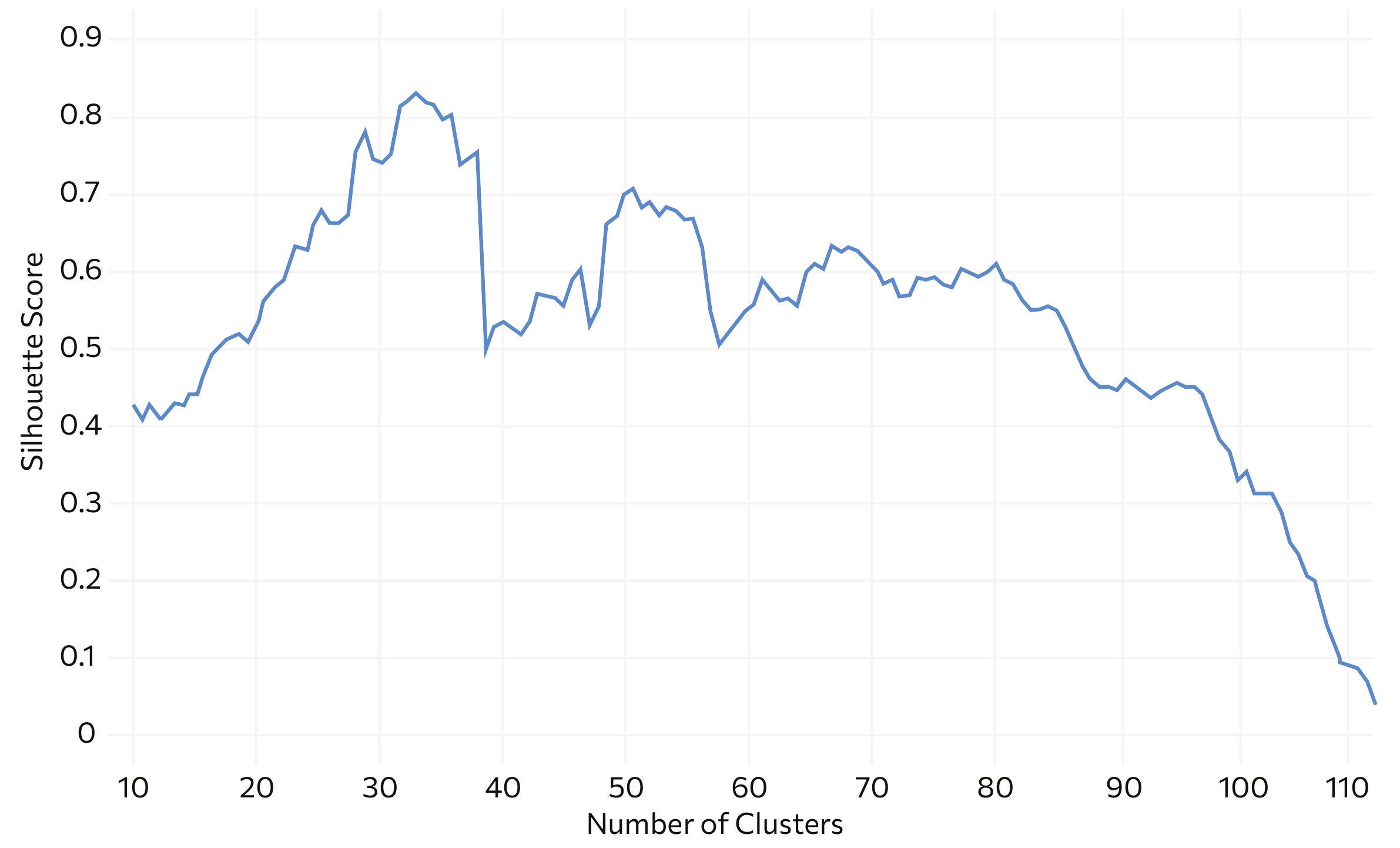}
\caption{The silhouette score of the clustering.}\label{fig:silhouette}
\end{figure}

Upon assessing the silhouette scores, we identified the optimal number of 33 clusters, achieving a mean silhouette score of $0.83$. Having executed the clustering process, the resulting clusters showed noticeable patterns related to the impact of code modifications on the software quality metrics. The distribution of modifications across these clusters showcased a spectrum of effects on complexity, coupling, documentation and size. Importantly, the presence of modifications from multiple repositories within these clusters ensured that we avoided the bias of repetitive modifications within the same repository or file. This result validates the robustness of our clustering process and the subsequent evaluation, reinforcing the usefulness of the clusters in deciphering the impact of code modifications on software quality metrics.

Upon applying the defined evaluation criteria, we discovered a set of meaningful clusters that captured the essence of our study. These clusters fulfilled our requirements, having a silhouette score greater than the mean silhouette score of the analysis, i.e. $0.83$, possessing at least $5$ modifications, and enclosing modifications across more than one repository.

The number of clusters that derived from the silhouette analysis, along with the respective number of clusters that fulfilled our criteria and the final silhouette score are depicted in Table \ref{tab:num_clusters}.

\begin{table}[ht]
\caption{The clustering statistics after evaluation.}\label{tab:num_clusters}
\centering
\begin{tabular}{lc}
\toprule
\textbf{Metric} & \textbf{Value}\\
\midrule
Number of Clusters & 33\\
Number of Clusters after Evaluation & 13\\
Average Silhouette Score & 0.83\\
\bottomrule
\end{tabular}
\end{table}

The analyzed clusters provided rich insights into the trends and patterns of code modifications and their impact on software quality. In this section, we delve into the analysis of some of these clusters (with IDs $10$, $27$ and $30$), produced by our comprehensive software quality assessment. Each cluster represents a distinct grouping of codebase modifications, shedding light on specific patterns and trends in our dataset.

In Table \ref{tab:cluster_10}, we present an in-depth look at the data within the cluster 10, showcasing the summary, as well as the simple summary, for each individual modification contained within it. This granular view allows us to gain insights into the characteristics and nature of the modifications comprising the cluster.

\begin{table}[H]
\caption{The modifications-entries of cluster 10.}\label{tab:cluster_10}
\centering
\begin{tabular}{cp{2.5in}p{2in}}
\toprule
Index & Summary & Simple Summary\\
\midrule
1 & Updated directory names and file paths in MediaCodec & Revised Directory Structure\\
\midrule
2 & Added 'Golden Sun' and 'Secret of Evermore' to the preset list. & Updated Preset List\\
\midrule
3 & Added "Freedom Planet" to game presets and submenu items & Updated game presets\\
\midrule
4 & Added "Clash at Demonhead" to list of game presets & Updated Game Presets\\
\midrule
5 & Added "RygarNes" to list and preset map. & Updated list and map.\\
\midrule
6 & Added "Shovel Knight" to game list and submenu map & Updated game list\\
\midrule
7 & Added "Stardew Valley" to game list and submenu map & Updated game list\\
\midrule
8 & Added "ShiningForce" to game list and preset map submenu. & Updated game list\\
\midrule
9 & Updated comments for better clarity in Activity class. & Improved comment clarity.\\
\bottomrule
\end{tabular}
\end{table}

Moving on to Table \ref{tab:cluster_10_metrics}, we shift our focus to the collective impact of all modifications within this cluster on static quality metrics. Here, we present the average influence that these modifications exert on a range of key quality metrics, offering a consolidated perspective on the alterations induced by the cluster as a whole. It should be noted that the colors in Table \ref{tab:cluster_10_metrics} indicate a positive (green) or negative (red) impact of the metric difference in the total quality of the codebase. For example, the increase of \emph{MI} (Maintainability Index) in the modifications of the cluster signifies a potential increase in code maintainability, as reflected in the positive (green) coloration in the table. This comprehensive analysis allows us to better understand the overarching effects of Cluster $10$ on the software's quality attributes and facilitates data-driven decisions for codebase management and optimization.

\begin{table}[H]
\footnotesize
\caption{The average impact of cluster 10 to quality metrics. \textit{Doc} stands for Documentation.}\label{tab:cluster_10_metrics}
\centering
\begin{tabular}{lclcc}
\toprule
 & \multicolumn{2}{c}{\textbf{Static Analysis Metrics}} & \multicolumn{2}{c}{\textbf{Computation Levels}}\\
\midrule
\textbf{Property} & \textbf{Name} & \textbf{Description} & \textbf{Method} & \textbf{Class}\\
\midrule
\multirow{12}{*}{\textbf{\rotatebox[origin=c]{90}{Complexity}}} & HCPL & Halstead Calculated Program Length & \textcolor{LimeGreen}{-32.0\%} & -\\
 & HDIF & Halstead Difficulty & \textcolor{LimeGreen}{-20.8\%} & -\\
 & HEFF & Halstead Effort & \textcolor{LimeGreen}{-37.1\%} & -\\
 & HPL & Halstead Program Length & \textcolor{LimeGreen}{-33.1\%} & -\\
 & HPV & Halstead Program Vocabulary & \textcolor{LimeGreen}{-23.4\%} & -\\
 & HTRP & Halstead Time Required to Program & \textcolor{LimeGreen}{-37.2\%} & -\\
 & HVOL & Halstead Volume & \textcolor{LimeGreen}{-36.4\%} & -\\
 & MI & Maintainability Index & \textcolor{LimeGreen}{+2.9\%} & -\\
 & McCC & McCabe's Cyclomatic Complexity & 0.0\% & -\\
 & NL & Nesting Level & 0.0\% & 0.0\%\\
 & NLE & Nesting Level Else-If & 0.0\% & 0.0\%\\
 & WMC & Weighted Methods per Class & - & 0.0\%\\
\midrule
\multirow{5}{*}{\textbf{\rotatebox[origin=c]{90}{Coupling}}} & CBO & Coupling Between Object classes & - & 0.0\%\\
 & CBOI & CBO Inverse & - & 0.0\%\\
 & NII & Number of Incoming Invocations & 0.0\% & 0.0\%\\
 & NOI & Number of Outgoing Invocations & 0.0\% & 0.0\%\\
 & RFC & Response set For Class & - & 0.0\%\\
\midrule
\multirow{4}{*}{\textbf{\rotatebox[origin=c]{90}{Doc}}} & AD & API Documentation & - & 0.0\%\\
 & CD & Comment Density & 0.0\% & 0.0\%\\
 & CLOC & Comment Lines of Code & 0.0\% & 0.0\%\\
 & DLOC & Documentation Lines of Code & 0.0\% & 0.0\%\\
\midrule
\multirow{2}{*}{\textbf{\rotatebox[origin=c]{90}{Size}}} & LLOC & Logical Lines of Code & \textcolor{LimeGreen}{-11.8\%} & 0.0\%\\
 & LOC & Lines of Code & \textcolor{LimeGreen}{-12.5\%} & 0.0\%\\
\bottomrule
\end{tabular}
\end{table}

Cluster 10 comprises a collection of codebase modifications, each with its own unique set of changes aimed at enhancing or optimizing the software. These modifications provide valuable insights into the diverse nature of alterations made during the development process. Notably, the cluster includes a range of activities such as updates and additions to various game-related lists and presets, as well as improvements to comment clarity in the Activity class. Analyzing the average impact of Cluster 10 on static quality metrics provides further insights into the consequences of these modifications:

\begin{itemize}
    \item Within the cluster, significant improvements in various complexity-related metrics are observed, indicated by the green color. Metrics such as HCPL, HDIF, HEFF, HPL, HPV, HTRP, and HVOL all show substantial reductions, suggesting a decrease in code complexity and effort required for programming.

    \item The Maintainability Index (MI) sees a slight increase, indicating an improvement in code maintainability.

    \item The size-related metrics, LLOC and LOC, both exhibit notable decreases, which could signify code optimization or removal of redundant lines.

    \item Other metrics related to coupling and documentation remain largely unaffected.
\end{itemize}

Based on the summaries provided in the first table, it can be inferred that the modifications within cluster 10 primarily involve updates and additions to game presets, along with some changes related to directory structure and comment clarity. Consequently, these modifications likely triggered a refactoring process aimed at improving code organization and readability, which contributed to reductions in complexity metrics such as Halstead Calculated Program Length, Halstead Difficulty, Halstead Effort, Halstead Program Length, Halstead Program Vocabulary, Halstead Time Required to Program, and Halstead Volume. Additionally, the refactoring efforts may have resulted in a decrease in the size of the codebase, as indicated by reductions in both Lines of Code (LOC) and Logical Lines of Code (LLOC). Overall, the modifications within cluster 10 appear to have positively impacted code quality by streamlining the codebase and enhancing its maintainability.

Table \ref{tab:cluster_27} presents an overview of cluster 27, including summaries of each modification. This table offers insights into the nature of the modifications within this cluster.

\begin{table}[H]
\caption{The modifications-entries of cluster 27.}\label{tab:cluster_27}
\centering
\begin{tabular}{cp{2.5in}p{2in}}
\toprule
Index & Summary & Simple Summary\\
\midrule
1 & Removed menu creation and item selection methods & Deleted functions\\
\midrule
2 & Implemented recursive approach for WordDictionary class & Implemented recursive programming\\
\midrule
3 & Refactored package name and updated parameters for spectrum audio processing & Refactored Package, Updated Parameters\\
\midrule
4 & Updated menu items in mnFileMenu with ActionListener methods & Implemented ActionListener methods\\
\midrule
5 & Updated build number and added 'Open Hex' option in File menu & Modified programmer settings\\
\midrule
6 & Fixed UI thread tests for API 28 and refactored animation verification for autofocus functionality. & Improved Animation Testing\\
\midrule
7 & Updated setUp method parameters and replaced SCREEN\_WINDOW\_NORMAL with SCREEN\_NORMAL & Adjusted method parameters\\
\bottomrule
\end{tabular}
\end{table}

In Table \ref{tab:cluster_27_metrics}, we analyze the collective impact of cluster 27 on static quality metrics. This table provides an average overview of how these modifications affect key quality metrics.

\begin{table}[H]
\footnotesize
\caption{The average impact of cluster 27 to quality metrics. \textit{Doc} stands for Documantation.}\label{tab:cluster_27_metrics}
\centering
\begin{tabular}{lclcc}
\toprule
 & \multicolumn{2}{c}{\textbf{Static Analysis Metrics}} & \multicolumn{2}{c}{\textbf{Computation Levels}}\\
\midrule
\textbf{Property} & \textbf{Name} & \textbf{Description} & \textbf{Method} & \textbf{Class}\\
\midrule
\multirow{12}{*}{\textbf{\rotatebox[origin=c]{90}{Complexity}}} & HCPL & Halstead Calculated Program Length & \textcolor{red}{+8.4\%} & -\\
 & HDIF & Halstead Difficulty & \textcolor{red}{+9.7\%} & -\\
 & HEFF & Halstead Effort & \textcolor{red}{+25.9\%} & -\\
 & HPL & Halstead Program Length & \textcolor{red}{+8.9\%} & -\\
 & HPV & Halstead Program Vocabulary & \textcolor{red}{+5.7\%} & -\\
 & HTRP & Halstead Time Required to Program & \textcolor{red}{+25.9\%} & -\\
 & HVOL & Halstead Volume & \textcolor{red}{+10.8\%} & -\\
 & MI & Maintainability Index & \textcolor{red}{-1.2\%} & -\\
 & McCC & McCabe's Cyclomatic Complexity & \textcolor{red}{+16.6\%} & -\\
 & NL & Nesting Level & 0.0\% & \textcolor{red}{+33.3\%}\\
 & NLE & Nesting Level Else-If & 0.0\% & \textcolor{red}{+22.3\%}\\
 & WMC & Weighted Methods per Class & - & 0.0\%\\
\midrule
\multirow{5}{*}{\textbf{\rotatebox[origin=c]{90}{Coupling}}} & CBO & Coupling Between Object classes & - & 0.0\%\\
 & CBOI & CBO Inverse & - & 0.0\%\\
 & NII & Number of Incoming Invocations & 0.0\% & 0.0\%\\
 & NOI & Number of Outgoing Invocations & 0.0\% & \textcolor{red}{+25.0\%}\\
 & RFC & Response set For Class & - & 0.0\%\\
\midrule
\multirow{4}{*}{\textbf{\rotatebox[origin=c]{90}{Doc}}} & AD & API Documentation & - & \textcolor{LimeGreen}{+20.0\%}\\
 & CD & Comment Density & 0.0\% & \textcolor{red}{-6.9\%}\\
 & CLOC & Comment Lines of Code & 0.0\% & 0.0\%\\
 & DLOC & Documentation Lines of Code & 0.0\% & 0.0\%\\
\midrule
\multirow{2}{*}{\textbf{\rotatebox[origin=c]{90}{Size}}} & LLOC & Logical Lines of Code & 0.0\% & \textcolor{red}{+8.33\%}\\
 & LOC & Lines of Code & 0.0\% & \textcolor{LimeGreen}{-8.2\%}\\
\bottomrule
\end{tabular}
\end{table}

Cluster 27 comprises a collection of codebase modifications, such as code refactoring, method implementation, user interface (UI) enhancements, and bug fixes. Examining the mean influence of Cluster 27 on static quality metrics offers deeper insights into the outcomes of these alterations:

\begin{itemize}
    \item Complexity Metrics: Several complexity metrics, including HCPL, HDIF, HEFF, HPL, HPV, HTRP, and HVOL, show an increase, indicating that the modifications have introduced more intricate code structures. This suggests a potential need for careful documentation and maintenance.

    \item Maintainability Index (MI) experiences a slight decrease, suggesting a minor reduction in code maintainability. This warrants attention to ensure continued ease of maintenance.

    \item McCabe's Cyclomatic Complexity (McCC) shows a substantial increase, indicating a higher degree of code complexity. It is essential to monitor this metric to prevent code from becoming overly convoluted.

    \item Nesting Levels (NL and NLE) in class instances have increased significantly, while class size metrics like Weighted Methods per Class (WMC) and Coupling metrics remain relatively unchanged.

    \item Documentation Metrics: API Documentation (AD) increases for classes, indicating improved documentation practices. Comment Density (CD) sees a minor decrease, possibly due to code refactoring.

    \item Size Metrics: Logical Lines of Code (LLOC) exhibit an increase, reflecting more complex logic in methods.

    \item In contrast, Lines of Code (LOC) decrease, suggesting code optimization and reduction in redundancy.
\end{itemize}

Based on the summaries provided in the first table for cluster 27, which involve various modifications such as removing methods, implementing recursive approaches, refactoring package names, updating parameters, fixing UI thread tests, and adjusting method parameters, it can be inferred that these modifications likely resulted in significant changes to the codebase. Increases in Halstead metrics (HCPL, HDIF, HEFF, HPL, HPV, HTRP, HVOL) and McCabe's Cyclomatic Complexity (McCC) suggest that the modifications introduced complexities in the codebase, potentially due to the implementation of recursive approaches and the addition of new functionalities. The increase in nesting levels (NL, NLE) also indicates a higher level of code complexity, possibly resulting from the implementation of more intricate logic. Finally, the decrease in Comment Density (CD) suggests a reduction in the density of comments relative to code, which could be attributed to the removal of some methods or the consolidation of functionality, while the increase in Logical Lines of Code (LLOC) suggests an overall increase in the size of the codebase, likely due to the addition of new functionalities or adjustments made to existing code.

Table \ref{tab:cluster_30} presents an overview of cluster 30, providing summaries for each modification within this cluster, offering valuable insights into the nature of these changes.

\begin{table}[H]
\caption{The modifications-entries of cluster 30.}\label{tab:cluster_30}
\centering
\begin{tabular}{cp{2.5in}p{2in}}
\toprule
Index & Summary & Simple Summary\\
\midrule
1 & Refactor tree traversal methods by removing redundant set operations & Optimize Tree Traversal Methods\\
\midrule
2 & Refined code formatting and set TagReader to final & Improved and finalized code\\
\midrule
3 & Refactored Camera test cases and added additional tests for supported video sizes & Refactoring and Testing Code\\
\midrule
4 & Increased lock wait time from 1500 to 3000 & Extended lock wait time\\
\midrule
5 & Refactored methods to auto-fetch package name for account registration & Refactored method fetching\\
\midrule
6 & Updated waiting times in wait calls & Modified times\\
\midrule
7 & Updated test cases and modified mocking behavior for Application class & Revised testing procedures\\
\midrule
8 & Refactor tile handling in map, implement category fetching for tiles and adjust object ID manipulation methods & Refactoring and Implementing Updates\\
\midrule
9 & Refactored and simplified method calls in Spotify and YouTube services. Removed unnecessary context references. Simplified command contributions and descriptions. Updated import paths. & Refactored and simplified code\\
\bottomrule
\end{tabular}
\end{table}

In Table \ref{tab:cluster_30_metrics}, we analyze the collective impact of cluster 30 on static quality metrics. This table provides an average overview of how these modifications affect key quality metrics.

\begin{table}[H]
\footnotesize
\caption{The average impact of cluster 30 to quality metrics. \textit{Doc} stands for Documentation.}\label{tab:cluster_30_metrics}
\centering
\begin{tabular}{lclcc}
\toprule
 & \multicolumn{2}{c}{\textbf{Static Analysis Metrics}} & \multicolumn{2}{c}{\textbf{Computation Levels}}\\
\midrule
\textbf{Property} & \textbf{Name} & \textbf{Description} & \textbf{Method} & \textbf{Class}\\
\midrule
\multirow{12}{*}{\textbf{\rotatebox[origin=c]{90}{Complexity}}} & HCPL & Halstead Calculated Program Length & \textcolor{LimeGreen}{-5.5\%} & -\\
 & HDIF & Halstead Difficulty & \textcolor{red}{+3.7\%} & -\\
 & HEFF & Halstead Effort & \textcolor{red}{+3.9\%} & -\\
 & HPL & Halstead Program Length & \textcolor{red}{+1.5\%} & -\\
 & HPV & Halstead Program Vocabulary & \textcolor{LimeGreen}{-4.3\%} & -\\
 & HTRP & Halstead Time Required to Program & \textcolor{red}{+3.9\%} & -\\
 & HVOL & Halstead Volume & \textcolor{red}{+0.1\%} & -\\
 & MI & Maintainability Index & \textcolor{red}{-2.2\%} & -\\
 & McCC & McCabe's Cyclomatic Complexity & \textcolor{red}{+25.0\%} & -\\
 & NL & Nesting Level & \textcolor{red}{+25.0\%} & 0.0\%\\
 & NLE & Nesting Level Else-If & \textcolor{red}{+25.0\%} & 0.0\%\\
 & WMC & Weighted Methods per Class & - & \textcolor{LimeGreen}{-35.0\%}\\
\midrule
\multirow{5}{*}{\textbf{\rotatebox[origin=c]{90}{Coupling}}} & CBO & Coupling Between Object classes & - & \textcolor{LimeGreen}{-50.0\%}\\
 & CBOI & CBO Inverse & - & 0.0\%\\
 & NII & Number of Incoming Invocations & 0.0\% & 0.0\%\\
 & NOI & Number of Outgoing Invocations & \textcolor{LimeGreen}{-8.3\%} & \textcolor{LimeGreen}{-50.0\%}\\
 & RFC & Response set For Class & - & \textcolor{LimeGreen}{-34.8\%}\\
\midrule
\multirow{4}{*}{\textbf{\rotatebox[origin=c]{90}{Doc}}} & AD & API Documentation & - & \textcolor{LimeGreen}{+20.0\%}\\
 & CD & Comment Density & 0.0\% & \textcolor{LimeGreen}{+16.6\%}\\
 & CLOC & Comment Lines of Code & 0.0\% & \textcolor{red}{-17.8\%}\\
 & DLOC & Documentation Lines of Code & 0.0\% & \textcolor{red}{-18.3\%}\\
\midrule
\multirow{2}{*}{\textbf{\rotatebox[origin=c]{90}{Size}}} & LLOC & Logical Lines of Code & \textcolor{red}{+12.0\%} & \textcolor{LimeGreen}{-30.5\%}\\
 & LOC & Lines of Code & \textcolor{red}{+13.0\%} & \textcolor{LimeGreen}{-28.2\%}\\
\bottomrule
\end{tabular}
\end{table}

The modifications within Cluster 30 exhibit several noteworthy trends and implications. Firstly, the optimizations and code refactoring efforts, as indicated by "Optimize Tree Traversal Methods", "Improved and finalized code", and "Refactoring and Testing Code", contribute to a reduction in complexity metrics such as Halstead Calculated Program Length (HCPL), Halstead Program Vocabulary (HPV), and Weighted Methods per Class (WMC). This signifies a positive impact on code maintainability, denoted by the green coloration in these metrics.

However, it is worth noting that certain modifications introduce complexity, such as an increase in McCabe's Cyclomatic Complexity (McCC) and Nesting Levels (NL and NLE), as well as an elevation in the Halstead Program Length (HPL) and Halstead Time Required to Program (HTRP). These changes might potentially reduce code readability and comprehensibility, denoted by the red coloration in the metrics.

The reductions in coupling metrics, particularly Coupling Between Object classes (CBO) and Response set For Class (RFC), along with the increase in Comment Density (CD), signify a positive influence on code modularity and maintainability. 

Overall, Cluster 30 demonstrates a balance between code optimization and increased complexity. These modifications collectively contribute to enhanced code maintainability and modularity while introducing some intricacies in the complexity of the codebase.

In this analysis, we have presented three distinctive examples of clusters, each showcasing varying impacts of codebase modifications on software quality metrics. The first example, Cluster 10, predominantly exhibits positive influences on metrics, signifying codebase improvements and optimization. Conversely, Cluster 30 demonstrates modifications that introduce a mix of positive and negative impacts on metrics, emphasizing the importance of carefully weighing trade-offs in code changes. Lastly, Cluster 27 predominantly reflects negative influences on metrics, highlighting potential areas of concern in code maintainability and complexity. These examples serve as insightful illustrations of the multifaceted consequences of code modifications, further emphasizing the need for data-driven decision-making in software development.

It is important to note that additional examples of top-silhouette clusters, along with their respective analyses, can be explored online\footnote{\url{https://gist.github.com/karanikiotis/11956966dd3f41e139d4c026237366ba}}. These clusters collectively paint a comprehensive picture of the dynamics and implications of codebase modifications on software quality metrics.

\section{Discussion}\label{sec:discussion}

Our study offers a unique exploration into the impact of code modifications on software quality metrics, particularly focusing on complexity and documentation. The proposed bottom-up methodology enabled us to effectively group code modifications based on their impact on these quality metrics.

The results of our study validates the complex interaction between code modifications and software quality as defined by previous research and advances by observing substantial variability between different types of code modifications and their impact on quality metrics. Some clusters of modifications had a more distinct effect on certain metrics than others, underlining the complicated nature of software development.

Furthermore, we show that different clusters exhibit different trends and patterns. For instance, some clusters are characterized by modifications that predominantly improve a specific aspect of software quality, while others are defined by modifications that seem to decay certain quality aspects. These trends and patterns offer valuable insights that can aid in improving coding practices and software quality.

One key observation from our study is the significance of context in assessing the impact of code modifications on software quality. While certain types of modifications generally lead to improvements in specific quality metrics, the actual impact often depends on the specific context of the modification, such as the overall structure of the code or the specific functionality being modified.

Finally, our study illustrates the potential of AI, in our case, the GPT-4 model, to assist in understanding code modifications. The generalized generated summaries provide a succinct, comprehensible description of each modification, facilitating our analysis and interpretation of the data.

\section{Conclusions and Future Work}\label{sec:conclusions-and-future-work}

Our work aims to clarify the impact of code modifications on software quality metrics. We harnessed an innovative approach incorporating popular GitHub repositories, static quality analysis, natural language processing with an AI language model, and clustering methods. Our findings demonstrate the influence of specific types of code modifications on software quality. 

In summary, we argue that our research bridges the gap in understanding the relationship between code modifications types and software quality, setting the groundwork for future research and technological advancement in this area.

Future work may involve a more detailed, case-by-case analysis of code modifications, and an investigation of other factors that could influence the relationship between code modifications and software quality. Additionally, our approach could be applied to other programming languages or codebases, and potentially incorporate other software quality metrics not considered in this study. Finally, our research could serve as the basis of a broader approach, that will recommend specific code modifications to developers that seek for improvement in specific areas of software quality, by leveraging the clusters generated by our analysis.

Overall, we believe that our study contributes to a growing body of research aimed at understanding and improving software quality, and feel confident that our findings can serve as a valuable resource for developers and teams striving for higher-quality software. Alongside automated source code generation, targeted code quality improvement may prove an optimal way for software evolution.

\bibliographystyle{elsarticle-num} 
\bibliography{bibliography}

\end{document}